\documentclass[fleqn,10pt]{wlscirep}

\usepackage{amsmath, epsfig, mathrsfs, graphicx, amssymb, color}

\usepackage{hyperref}
\usepackage{array}
\usepackage{booktabs}
\usepackage{yquant}
\useyquantlanguage{qasm}
\usepackage{physics}
\usepackage{todonotes}  % TODO boxes
\usepackage{soul}  % Strikethrough
\usepackage{enumitem}  % Different enumerate env. labels
\usetikzlibrary{quotes}

\def\cor#1{{#1}}
\def\mb{\begin{pmatrix}}
	\def\me{\end{pmatrix}}
\def\be#1\ee{\begin{equation}#1\end{equation}}
\newcommand{\ba}{\begin{eqnarray} }
	\newcommand{\ea}{\end{eqnarray} }

\title{Cross-platform certification of the qubit space with a minimal number
		of parameters}

\author[1,2,3]{Tomasz Rybotycki}
\author[4]{Tomasz Bia{\l}ecki}
\author[5,6]{Josep Batle}
\author[4]{Jakub Tworzyd{\l}o}
\author[4,*]{Adam Bednorz}
\affil[1]{Systems Research Institute, Polish Academy of Sciences, 6 Newelska Street, PL01-447 Warsaw, Poland}
\affil[2]{ Nicolaus Copernicus Astronomical Center, Polish Academy of Sciences, 18 Bartycka Street, PL00-716 Warsaw, Poland }
\affil[3]{ Center of Excellence in Artificial Intelligence, AGH University, 30 Mickiewicza Lane, PL30-059 Cracow, Poland }
\affil[4]{Faculty of Physics, University of Warsaw, ul. Pasteura 5, PL02-093 Warsaw, Poland}
\affil[5]{Departament de F\'isica and Institut d'Aplicacions Computacionals de Codi Comunitari (IAC3), Campus UIB, E-07122 Palma de Mallorca, Balearic Islands, Spain}
\affil[6]{CRISP -- Centre de Recerca Independent de sa Pobla, 07420 sa Pobla, Balearic Islands, Spain}

\affil[*]{Adam.Bednorz@fuw.edu.pl}

	\begin{abstract}
		We demonstrate a determinant dimension witness of a qubit space. Our test
		has a minimal number of independent parameters. We achieve it by mapping the
		Bloch sphere $\pi/2$-rotation axis angle on the non-planar so-called Viviani
		curve. We ran our test on different platforms: IBM Quantum, IQM Resonance, and
		IonQ. Our investigations show that numerous qubits, especially from the newest IBM Heron family devices, fail the test by more than ten standard deviations. The nature of
		those deviations has no simple explanation as the test is robust against common
		imperfections.
	\end{abstract}
	
\begin{document}
\flushbottom	
	\maketitle
	\thispagestyle{empty}
	
	\section*{Introduction}	
	Quantum computers rely on two-level systems, commonly known as qubits. Computations
	require the qubit space to be reliable, not combined with a larger space in particular. The
	limit on the computational space dimension is critical for a fault-tolerant quantum
	computation performed on real devices, as the error correction relies on the restricted
	model of noise \cite{mit1,mit2,mit3}.
	
	Quantum space dimension can be verified by a dimension witness
	\cite{gallego,hendr,ahr,ahr2,dimbell,dim1}. Such witnesses implement a two-stage
	protocol -- the initial preparation and the final measurement, both chosen independently.
	A special case of dimension witnesses are null witnesses \cite{dim}. For a system
	below a specified dimension, the value of such witness is -- up to statistical
	error -- zero, and nonzero otherwise. In a sense, a null witness certifies the
	linear independence of the outcome probability $p(M|P)$ for the preparation $P$
	and measurement $M$ \cite{dim,chen,bb22,opt}.
	
	Dimensionality tests are robust against several errors of existing quantum gates.
	Microwave pulses, used to perform physical operations on the qubits, suffer
	from nonlinearities of waveform generators \cite{distor}. Therefore, a simple
	deviation of the probability distribution from the theoretical prediction
	\cite{praca} is not a sufficient certification of an extra qubit space. We have
	already performed such tests on IBM Quantum \cite{epj}, and its variant for a
	single repeated gate \cite{pra}, and found significant deviations.
	
	In the present contribution, we simplify the witness further
	\cite{bb22}, with just two independent parameters (angles) to control the preparation
	and measurement. Such approach minimizes the possible correlations between the
	preparation and measurement parameters. We tested a set of qubits at various times,
	on  multiple platforms: IBM Quantum, IonQ, and IQM Resonance. The main advantages of this test
	are:
	\begin{itemize}
		\item high precision,
		
		\item minimal preparation-measurement correlation,
		
		\item no two-qubit gates,
		
		\item robustness against relaxation and dephasing errors,
		
		\item cross-platform compatibility,
		
		\item reproducibility.
	\end{itemize}

	Some systems passed the experimental tests, and for them the two-level
	qubit model holds. However, there are many cases  failing the test at 5--10 and more standard
	deviations,  the newest IBM Heron computers in particular. Although test results vary in time,  for IBM Heron family devices the deviations are generally consistent. Some qubits tested a month later
	passed the test, while the others failed. Moreover, the deviations on IonQ devices
	occur incidentally. This result shows that one cannot detect the cause in the routine
	calibration. Whatever the reason, it needs an urgent explanation to advance efforts
	of efficient error correction.
	
	\section*{Description of the test}
	
	The test is applied to the qubit space $d=2$ with the witness constructed from
	the measurement probabilities for the state $P$ and the measurement operator
	$M$, associated with one of the outcomes, say $0$. According to quantum theory,
	the probability reads $p(M|P)=\mathrm{tr}M P$, where the preparation operator is
	$P=P^{\dag}\geq 0$, $\mathrm{tr}P=1$ and the measurement is
	$1\geq M=M^{\dag}\geq 0$. 
	
	We propose a determinant dimension witness test based on a $5 \times 5$ matrix of probabilities $F$. The entries of $F$ are $F_{kj}=p(M_{j}|P_{k})$ for $k = 1, 2, ..., 5$ and $j = 1, 2, 3, 4.$ Since we need 5 parameter values for state preparation but only 4 values for the measurement, we need to supplement the final row with ones, formally $F_{5j}=1$. This way we still expect determinant witness $W := \det F$ to be 0 in the ideal case, that is if all $P_{j}$ and $M_{k}$ represent the same two-level space \cite{leak, bb22}. This is a straightforward consequence linearity of the trace with respect to the matrices, which have maximally $d^2$ entries in the quanum space of $d$ dimensions. Exceeding this limit makes columns/row linearly dependent and the determinant must vanish.
	
	Moreover, the determinant remains zero even if all preparations contain some
	constant incoherent leakage term, namely $P'_{j}=P_{j}+\tilde{P}$, with
	$\tilde{P}$ in an arbitrary	space, independent of $j$. In this sense, the
	uncontrolled leakage to an extra space does not affect the test
	\cite{leak,gam}.
	
	If the quantum system consists of only two levels (like qubit), we expect
	$W=0$. Any deviation should remain within the statistical error due to
	finite statistics. Otherwise, the deviation would indicate that the actual
	space is larger. Assuming a two-dimensional Hilbert space, we construct the
	states and	the operators in the computational basis $\ket{0}$, $\ket{1}$.
	We also assume the initial state is $\ketbra{0}{0}$.	
	Additionally, the test relies on the following assumptions:
		\begin{itemize}
			\item independent and identical probability distributions of all repetitions,
			\item independence between the  preparation and measurement operations.
	\end{itemize}
	
	Since we aimed to run our test on a real devices, we had to decide what
	gates to use in our test. It would benefit the experiment greatly, if the
	gates were native to the target device(s), since then the gate-induced
	error would be the lowest. Previously, we tested IBM Quantum devices \cite{praca, pra, epj}, so
	we had the most experience with them. We decided to use the
	devices' native gates,  i.e. use the simplest steering pulse sequences when constructing the quantum circuits, not only for IBM but also IQM and IonQ.
    The native gate realized by IBM Quantum, IQM and IonQ is the $\pi/2$ rotation on the Bloch
	sphere in $\{\ket{0}$, $\ket{1}\}$ space:
	\begin{equation}
		S=\frac{1}{\sqrt{2}}
		\begin{pmatrix}
			1  & -i \\
			-i & 1
		\end{pmatrix}\label{smat}.
	\end{equation}
	Moreover, the rotation axis can itself be rotated by a given angle $\theta$. This rotation
	is realized by an $S$ gate and two auxiliary $Z(\theta)$ (phase shift) gates,
	\begin{equation}
		S_{\theta}=Z^{\dag}_{\theta} SZ_{\theta} ,\: Z_{\theta}=
		\begin{pmatrix}
			e^{-i\theta/2} & 0             \\
			0              & e^{i\theta/2}
		\end{pmatrix}.
	\end{equation}

	The physical experiment describing our test is a sequence of qubit state $\ket{0}$
	initialization, followed by $S$, $S_{\beta}$ gates (preparation), then the two
	gates $S_{\phi}$, $S$ (measurement), and the readout pulse for the measurement
	of the state $\ket{0}$ again, see Fig. \ref{gates}. The parameters are $5$ preparation
	angles $\beta_{j}$ ($j=1..5$) to be chosen independently of the $4$ measurement
	angles $\phi_{k}$ ($k=1..4$). It gives the prepared state
	$P_{\beta}=S_{\beta} S\ketbra{0}{0}S^{\dag} S^{\dag}_{\beta}$ and the
	measured state
	$M_{ \phi}=S^{\dag}_{\phi} S^{\dag}\ketbra{0}{0}S S_{\phi}$, in terms of
	operators. We also have to assume that the gates/operations do not depend on
	the previous ones. To sum up, we can only test the combination of assumptions:
	dimension of the space and independence of operations.
	
	%%%%%%%%%%% Bloch vector description %%%%%%%%%%%%%%%%%%%
	One can use Bloch sphere representation of prepared states and measurements,
	using vectors $\boldsymbol n$ to represent the state
	$P=\ketbra{\boldsymbol p}{\boldsymbol p}=(1+\boldsymbol p\cdot\boldsymbol
	\sigma)/2$, with Pauli matrices $\sigma_{1,2,3}$. The initial state
	$\ketbra{0}{0}$ corresponds to the vector $(0,0,1)$. The $\pi/2$ rotations
	by $S$ gate and phase shifts $Z_{\theta}$ can be used to obtain the non-planar
	parametric Viviani curve on the Bloch sphere. Let us recall that the Viviani
	curve arises as the intersection of a cylinder tangent to a sphere and
	embedded inside it, passing through the origin. In our specific operation sequences, the
	vectors $\boldsymbol p_{\beta}$, $\boldsymbol m_{\phi}$, form respective opposite
	Viviani curves $-(\sin\beta\cos\beta,\sin^{2}\beta,\cos\beta)$ and $(\sin\phi\cos
	\phi,\sin^{2}\phi,\cos\phi)$. Viviani curves are especially useful in our test because:
	\begin{enumerate}[label=(\alph*)]
		
		\item it is non-planar, meaning it spans  complete
		qubit space, in contrast to simple circles,
		
		\item they are realized by a  native gates with single parameter ($\beta$ and $\phi$) minimizing the 
		unwanted correlations between the preparation and measurement operations,
		
		\item  the actual sets of angles are chosen to make the adjugate matrix $\mathrm{Adj}\:F$
		(contrary to inverse matrix, the adjugate one exists even if $W=0$)
		 distant from zero as possible to increase the test's sensitivity  to potential external states.
		
	\end{enumerate}
	The restriction to Viviani curve is not obligatory. One is free to choose any set of preparations and measurements,  similiary to \cite{epj}. However,  our goal was to construct a simple cross-platform qubit dimension test, hence the choice of Viviani curve angles.

\begin{figure}
		\includegraphics[scale=.4]{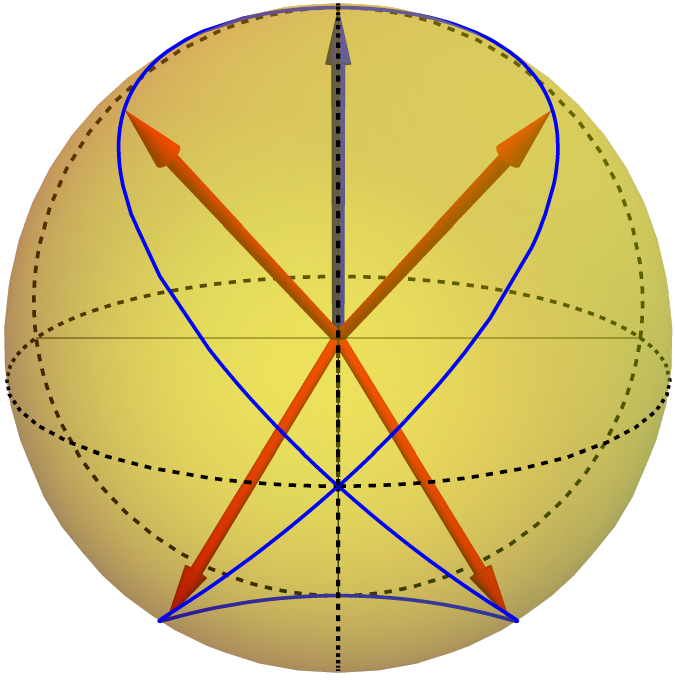}
		\caption{The Bloch vectors for the preparations $\boldsymbol p$ (red and blue arrows) and measurements $\boldsymbol m$ (red arrows) corresponding to the angles used in the	test, positioned on the Viviani curve (blue curve).}
		\label{bloch1}
	\end{figure}

	Then the probability matrix elements read
	\begin{equation}
		F_{kj}=\mathrm{Tr}M_{k}P_{j}=(1+\boldsymbol p_{j}\cdot\boldsymbol m_{k})/2.
	\end{equation}
	 We used
	$$\beta=\{\pi/4, -\pi/4, 3\pi/4, -3\pi/4, 0\},$$
	and $\phi_{k}=-\beta_{k}$ for $k=1\dots 4$. Respective Bloch vectors and their
	positions on the Viviani curve are presented in Fig. \ref{bloch1}.
	
	Theoretically, we can calculate the probability for every $\beta$ and $\phi$.
	Practically, the test is stochastic, and we have to estimate the error due to
	finite statistics. From Laplace expansion 
	\be 
	\delta W=\sum_{kj}\delta F_{kj}(\mathrm{Adj}\:F)_{jk}
	\ee
	where Adj is the adjugate matrix, and $\delta F= F-\langle F\rangle$ is the deviation of empirically obtained probability (frequencies) matrix $F$ from  it's theoretical form. The variance of $W$ is
	\be
	T\sigma^2=T\langle W^2\rangle\simeq \sum_{kj}F_{kj}(1-F_{kj})(\mathrm{Adj}\: F)_{jk}^2,
	\label{err}
	\ee
	where $T$ is the experiment repetitions number \cite{bb22}. The data we
	collected allow estimating the above error, which is  necessary for the
	results confidence level estimation / determination.
	
	\begin{figure}
		\centering
		\begin{tikzpicture}\scalebox{1.3}{
			\begin{yquant*}
				qubit {} q[1];
				init {$\ket 0$} q[0];
				
				[this subcircuit box style={dashed, "Preparation"}]
				subcircuit {
					qubit {} q[1];
					box {$S$} q[0];
					box {$S_{\beta}$} q[0];
				} (-);
				
				[this subcircuit box style={dashed, "Measurement"}]
				subcircuit {
					qubit {} q[1];
					box {$S_{\phi}$} q[0];
					box {$S$} q[0];
				} (-);
				
				measure q[0];
			\end{yquant*}}
		\end{tikzpicture}
		
		\caption{The quantum circuit for the dimension test with two parameters, angles
			$\beta$ and $\phi$. The protocol starts from the initial state $\ket{0}$, followed
			by preparation phase, gates $S$ and $S_{\beta}$ and the measurement phase, $S_{\phi}$ 
and $S$, with the final readout in the computational basis.}
		\label{gates}
	\end{figure}
	
	\begin{figure*}
		\includegraphics[scale=.7]{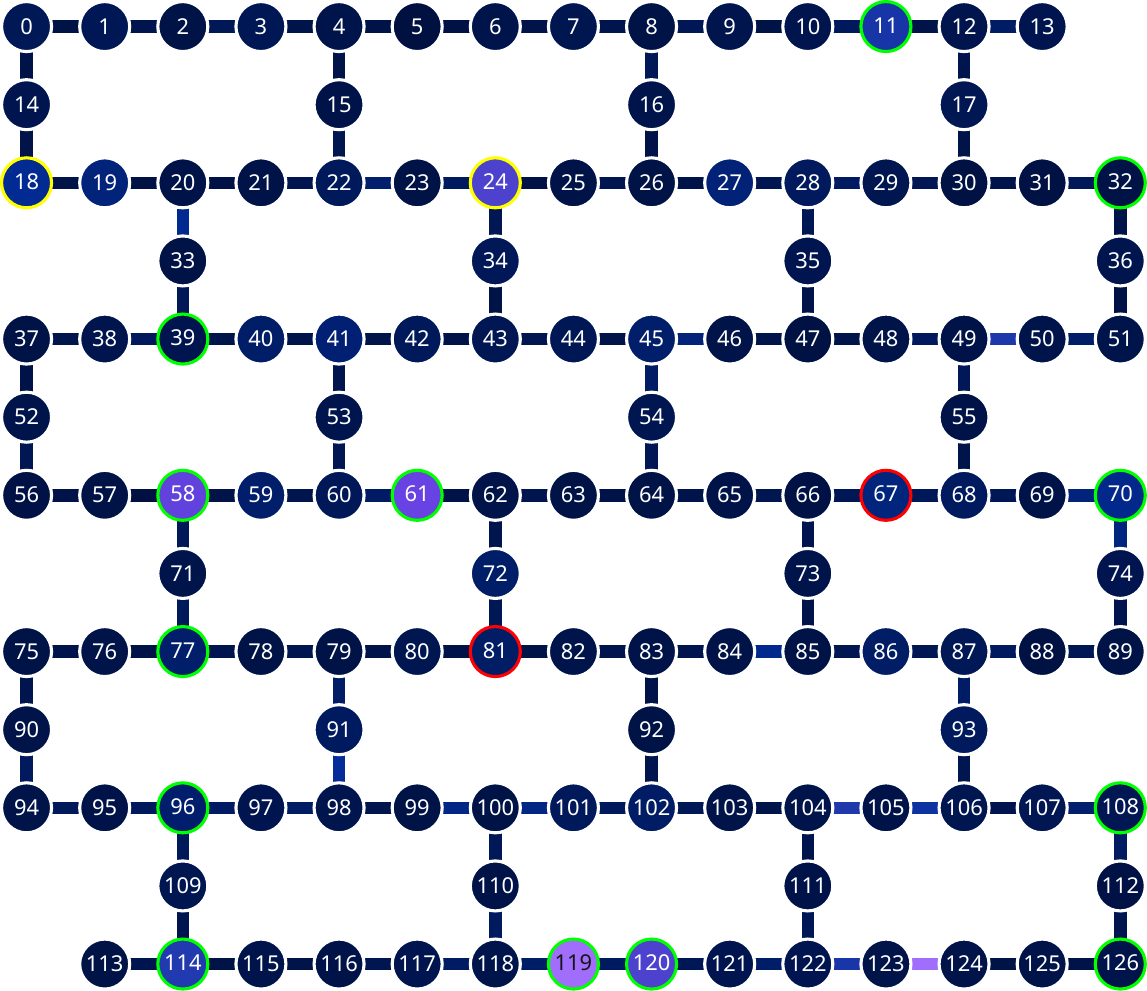}
		\caption{The topology of \texttt{ibm\_brisbane} device. We  highlighted the most faulty
			qubits tested in February -- yellow, and in March 2024 -- red. The rest
			of the qubits, highlighted green, have passed the test. Two-qubit Echoed Cross Resonance
			gates, unused in the test, connect the qubits.}
		\label{bri}
	\end{figure*}
	
	\begin{figure}
		\includegraphics[scale=.54]{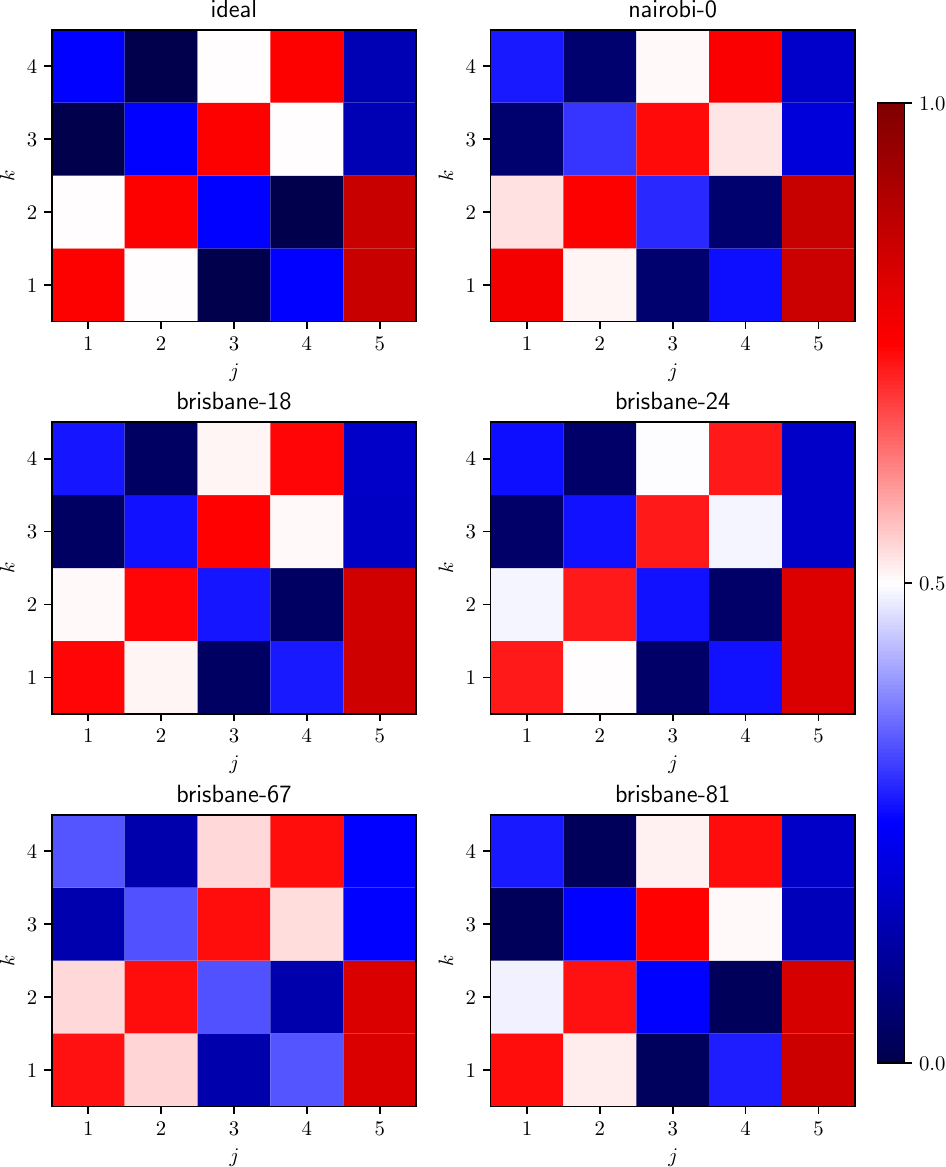}
		\caption{The probabilities $F_{kj}$ for the test using selected quantum states and measurement represented by Bloch vectors on the Viviani curve. From the top-left: ideal,
			\texttt{ibm\_nairobi} qubit 0, \texttt{ibm\_brisbane} qubits 18, 24 (Feb 2024), and 67, 81 (Mar
			2024).}
		\label{res}
	\end{figure}
	
	\begin{figure}
		\includegraphics[scale=.7]{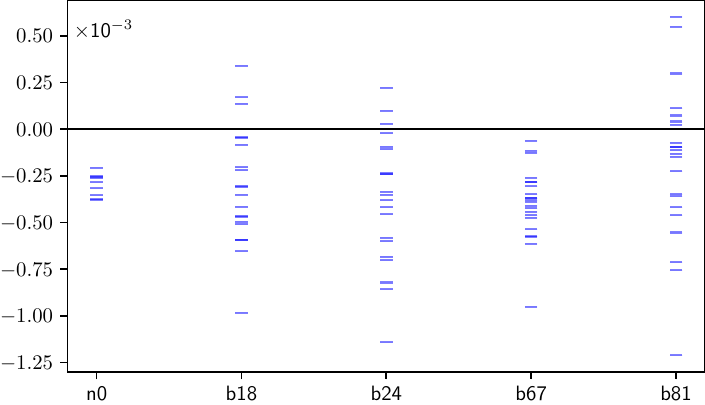}
		\caption{The values of the witness for individual jobs: \texttt{ibm\_nairobi} qubit 0,
			\texttt{ibm\_brisbane} qubits 18, 24 (Feb 2024) and 67, 81 (Mar 2024). We use symbols $n\#$ and $b\#$ for \texttt{ibm\_nairobi} and \texttt{ibm\_brisbane} qubits respectively.}
		\label{scat}
	\end{figure}

	\section*{Experiments}
	\label{sec:exp}
	
	We have run the above test on three publicly available quantum computing platforms:
	IBM, IQM, and IonQ. We used identical circuits and randomly shuffled the angles.
	Our test doesn't require a perfect implementation of gates. Standard local sources of depolarizing
	and relaxation errors, and general gate and readout errors do not affect our test,
	as they are assumed to remain within the two-level space. 
	Moreover, the test accounts for leakage to external states (e.g., $\ket{2}$)
	as long as it is incoherent and does not depend on the circuit parameters $\beta
	,\phi$. That is because occurence of such phenomenon adds a global constant to each $F_{jk}$ \cite{leak,gam}. {We present the comparison
		of single qubit gate properties for respective platforms in Table \ref{comm}.
	
	Due for nonlinearity of determinant and to avoid calibration dependence, we have calculated $W$ using data from IBM and IQM in two ways:
	\begin{enumerate}[label=(\roman*)]
		\item determining $F$ for each job (basic complete comutation task unit),  finding determinant $W_{\text{job}}$ for each job, and finally averaging over $W_{\text{job}}$,
		\item first averaging $F$ over all jobs and then  computing $W$. 
	\end{enumerate}
	The latter approach failed on IonQ because of large drifts in the data. We suspect varying calibrations were the cause of the drift.
	
	\begin{table}
		\begin{tabular}{*{6}c}
			\toprule platform & gate [s]          & measure [s]       & T1 [s]          & T2 [s]          & gate error       \\
			\midrule IBM      & $5\cdot 10^{-8}$ & $1\cdot 10^{-6}$ & $\sim 10^{-4}$ & $\sim 10^{-4}$ & $\sim 10^{-4}$   \\
			IQM               & -                & -                & $\sim 10^{-5}$ & $\sim 10^{-5}$ & $\sim 10^{-3}$   \\
			IonQ              & $\sim 10^{-4}$   & $5\cdot 10^{-5}$ & $10$           & $1.5$          & $5\cdot 10^{-4}$ \\
			\bottomrule
		\end{tabular}
		\caption{Comparison of qubit and gate properties for different platforms: single-qubit
			rotation gate time, qubit measurement time, relaxation decay time T1,
			decoherence time T2, single qubit gate error. The IBM families, Falcon (\texttt{ibm\_nairobi}), Eagle (\texttt{ibm\_brisbane}), Heron (\texttt{ibm\_torino}, \texttt{ibm\_pittsburgh}) differ rather by architecture and overall performance, while the single-qubit operations change within the similar order of magnitude.}
		\label{comm}
	\end{table}
	
	\begin{figure*}
		\includegraphics[scale=.7]{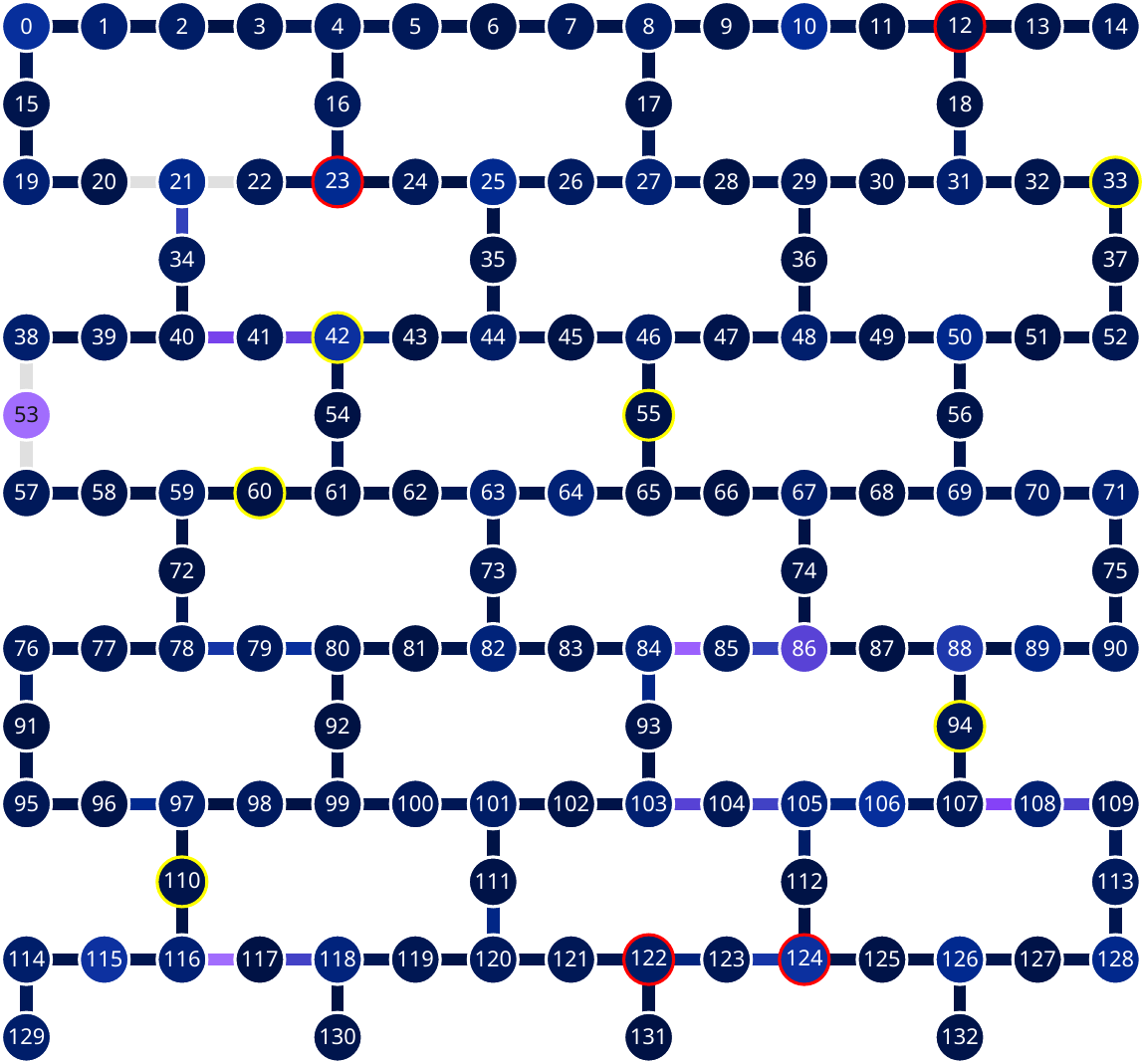}
		\caption{The topology of \texttt{ibm\_torino} device. We  highlighted the 
				qubits tested in September 2025 with yellow and red. The latter show exceptionally large violation.}
		\label{torm}
	\end{figure*}
	
	\begin{figure*}
		\includegraphics[scale=.7]{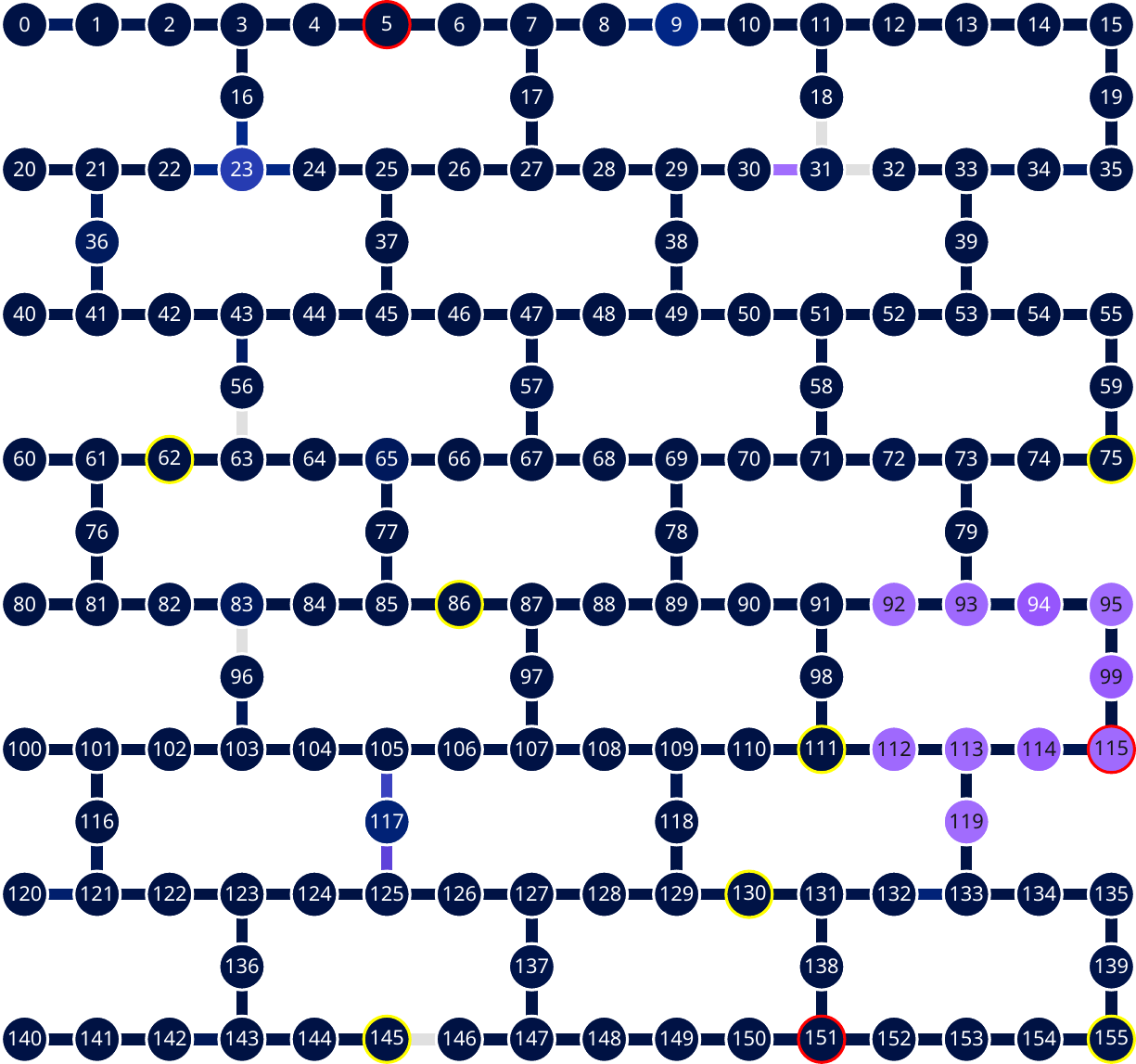}
		\caption{The topology of \texttt{ibm\_pittsburgh} device. We  highlighted the 
				qubits tested in September 2025 with yellow and red. The latter show exceptionally large violation.}
		\label{pitm}
	\end{figure*}
	
	The data and scripts used in the experiments are available at the public repository \cite{zen}.
	
	\subsection*{Hardware specification}
	
	The IBM Quantum, IonQ, and IQM Resonance platforms offer a list of devices consisting
	of many qubits. The qubits are controlled by a user-defined set of gates (operations) -- either
	single-qubit or two-qubit ones, parametrized by phase shifts.
	
	% Qubit realization 
	The IBM and IQM qubits
	are physical transmons \cite{gam2,transmon}, i.e. artificial quantum states
	existing due to the interplay of superconductivity (Josephson effect) and
	capacitance (Coulomb charging effect). The transmon consists in principle of a
	sequence of levels $\ket 0$, $\ket 1$, $\ket 2$, and so on, denoted shortly as
	$0,1,2,\dots$. For (contemporary) quantum computers to work correctly, we expect the devices
		to restrict their operations range to $\ket{0}$ and $\ket{1}$. On the other hand, the IonQ qubits
		are trapped ytterbium ions in the space spun by
		two hyperfine-split states \cite{ytt}. 
	
	% Gates realization
	Each gate is realized by a microwave pulse
	of the drive frequency (equivalent to energy difference between $0-1$ levels) for
	transmons or alternatively by Raman transitions for ions, \cite{qis}. The pulse frequency
	for transmons is in the range $4-5$ GHz, and $\sim 12$GHz for ions. It is worth mentioning,
		that the $Z_{\gamma}$ gate}
	is not actually a real pulse, but is realized by adding a rotation (phase shift)
	between real and imaginary components of the pulse for  upcoming gates
	\cite{zgates}. In the case of transmons, due to the anharmonicity (difference between $0-1$ and
	$1-2$ transition frequencies) of about 300MHz, the working space effectively collapses
	into two states $0$ and $1$.
	
	% Readout
	The
	readout of transmons is performed by coupling the resonator to another long microwave
	pulse at a frequency different from the drive frequency, and finally measuring the
	populated photons \cite{qis,read}. Ions are measured by laser pulse stimulating
	transitions to specific levels.
	
	\subsection*{IBM Simulations}
	
	To ensure our test works as intended, we ran it on the \texttt{qiskit aer} simulator. First
	we assumed the perfect gates and no noise. This test was done on a single qubit, because
	without the noise, simulator qubits are indistinguishable. We ran a single job with 
	$10^6$ shots. We got $W_{\mathrm{sim}} = (-13.8 \pm 9.9) \cdot 10^{-5}$.

\begin{figure}[ht!]
		\includegraphics[scale=.65]{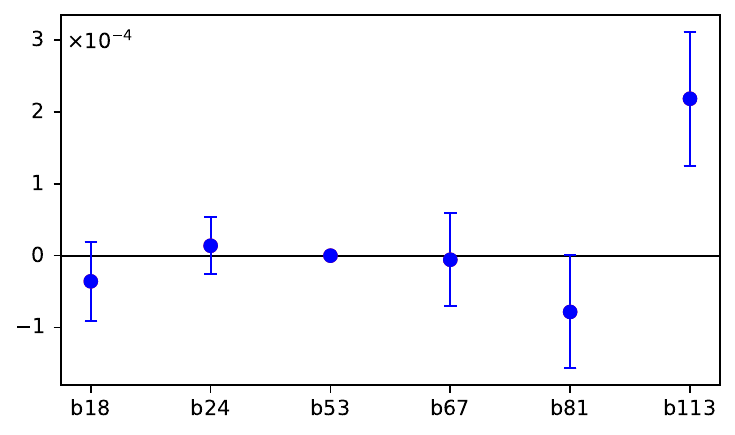}
		\caption{The value of the witness $W$ of the selected simulated \texttt{ibm\_brisbane}
			qubits denoted b$n$, with $n$ being the qubit number. Aside 
			from the qubits we used for the experiment on a real device, we also present the
			least (b53) and the most (b113) noisy qubit.}
		\label{fig:sim_noisy}
	\end{figure}

	We then repeated our tests, this time on a noisy simulator. We targeted each qubit of the
	\texttt{ibm\_brisbane} backend. Since the gate errors  vary from qubit to qubit, 
	we wanted to see if this variation will show up in the simulations. We ran a single job
	with $10^6$ shots for each qubit. The results are shown in the figure \ref{fig:sim_noisy}.
	
	We can see that even on the simulator the disproportion between qubits performance is 
	significant. While the least noisy qubit had the witness value of order $10^{-8}$, the value
	of $W$ for the most noisy one was of order $10^{-4}$. We have rerun the simulations in September 2025 imitating IBM Heron devices. In the latest simulations, no siginificant deviations have been found.
	
	\subsection*{IBM}
	
	% Devices desc	
	Initially, we ran the test on two IBM Quantum devices: \texttt{ibm\_nairobi} and \texttt{ibm\_brisbane} \cite{ibmq}.
	The first was a 7-qubit system on which we tested the qubit 0 in October 2023.
	The latter is a 127-qubit system. The declared coherence times are of the order
	$50\cdot 10^{-6}$~s, while the single-qubit gates take $\sim 50\cdot 10^{-9}$~s
	and readout $\sim 1.2\cdot 10^{-3}$~s. To use it efficiently, we have chosen two
	sets of 10 qubits, the first in February and the second in March 2024.
	
	% Running the experiments
	To run the experiments on IBM devices, one sends a list 
	of so-called jobs for execution.  Experiments can consist of
		multiple jobs, which can be repeated a number of times. Aside from defining quantum circuits to be executed on
		the target device, one has to specify additional job parameter. One of those, the number
	of shots, specifies the
	repetition number of the preprogrammed sequence of circuits within a single job.
	Note that each execution of a shot means running full sequence of circuits, then proceeding
	to the next shot. To avoid memory-related effects, we randomly shuffled the experiments
	within each job. The total number of each circuit executions is thus given by
	$T=\#\mathrm{jobs}\cdot \#\mathrm{shots}\cdot\#\mathrm{repetitions}$. The readout
	count for each circuit is the value returned after the job execution is complete.
	
	% Experiments desc
	For \texttt{ibm\_nairobi}, we ran nine jobs at $10^5$ shots and 15 repetitions. For
	\texttt{ibm\_brisbane}, we ran 20 jobs in February and 22 in March 2024, with $10^4$
	shots and ten repetitions each, because of access limitations enforced by IBM.
	We could measure ten qubits simultaneously, treating them as independent, because
	their positions on the qubits grid are isolated, see Fig. \ref{bri}. Moreover,
	the random shuffling of the parameters was applied to each qubit independently,
	so any crosstalk is unlikely.
	
	% Results
	In Tables
	\ref{data_b} and \ref{data_n}, we present the results and parameters of the qubits for both runs.
	The tests on \texttt{ibm\_nairobi} qubit 0 and \texttt{ibm\_brisbane} qubits -- 18, 24 in February and 67, 81 in March
	2024 -- resulted in the witness's value at five or more standard deviations. These
	cases are presented in detail in Fig. \ref{res} (the matrix of probabilities)
	and in Fig. \ref{scat} (results of individual jobs). 
	
	% New Heron results
	As of late, introduced  newer, Heron family devices: \texttt{ibm\_torino} with 133 qubits and \texttt{ibm\_pittsburgh}
		with 156 qubits. In September 2025, we have run analogous tests on both of them, with 40 and 30 jobs, respectively.
		We chose the qubits shown in Figs. \ref{torm} and \ref{pitm}.
		The tests showed significant violations  with negative determinant $W$ values, see Table \ref{data2}.
	
	% Additional notes.
	We stress that the gate error  of order $10^{-4}$ could not cause such deviations. That is because in such case the deviations should be of the order of the square of	the gate errors $\sim 10^{-8}$ (adding simultaneously to the preparations and measurements).
	The temporal change in the witness value  hints that the cause comes from the device calibrations,
	not just the pulse frequencies, which vary only slightly. However, consistent and large negative-valued violation 
		of the dimension witness in Heron devices indicate a
		serious operational issue that has to be investigated. 
The leakage to higher
	states, i.e., $\ket{2}$, is negligible \cite{leak,praca}, as its effects are much below the gate
	errors (see also our discussion in \cite{epj}).
	
	\begin{table*}
		\begin{tabular}{*{6}{c}}
			\toprule drive freq. (GHz) & gate error $10^{-4}$ & $W^{i}[10^{-5}]$ & $\sigma^{i}[10^{-5}]$ & $W^{ii}[10^{-5}]$ & $\sigma^{ii}[10^{-5}]$ \\
			\midrule 5.26              & 2.5                  & {\bf -29.8}      & {\bf 2.3}             & {\bf -29.9}       & {\bf 2.3}              \\
			\bottomrule
		\end{tabular}
		\caption{Experiment results for \texttt{ibm\_nairobi} qubit 0 (test in October 2023).
			The data contain drive frequency (inter-level), error of the $S$ gate, the witness $W^{i/ii}$ and the
			standard deviation $\sigma^{i/ii}$. Indices $i$ and $ii$ of $W$ and $\sigma$ indicate the
				$W$ calculation strategy discussed in section \ref{sec:exp}. Faulty qubits, \cor{with $W$ nonzero beyond 5 standard deviations,} are bolded.}
		\label{data_n}
	\end{table*}
	\begin{table*}
		\begin{tabular}{c|*{11}{c}}
			\toprule qubit (February 2024) & {\bf 18}  & {\bf 24}  & 32    & 39        & 58    & 61    & 70        & 114   & 119   & 126   \\
			\midrule drive freq. [GHz]     & 4.788     & 5.101     & 4.910 & 4.917     & 4.887 & 4.794 & 4.899     & 4.822 & 4.803 & 4.908 \\
			gate error $[10^{-4}]$         & 2.1       & 1.6       & 2.9   & 2.0       & 2.2   & 6.2   & 2.9       & 2.0   & 3.2   & 2.4   \\
			$W^{i}[10^{-5}]$               & {\bf -30} & {\bf -39} & -18   & -1.2      & 1.7   & -3.6  & -18       & -1.2  & -17   & 19    \\
			$\sigma^{i}[10^{-5}]$          & {\bf 6.1} & {\bf
				
				5.5}                           & 6.4       & 5.4       & 6.5   & 6.2       & 4.5   & 2.6   & 5.2       & 6.5                   \\
			$W^{ii}[10^{-5}]$              & {\bf -30} & {\bf
				
				-38}                           & -18       & -1.4      & 1.7   & -3.6      & -18   & -1.3  & -17       & 19                    \\
			$\sigma^{ii}[10^{-5}]$         & {\bf
				
				6.1}                           & {\bf5.5}  & 6.4       & 5.4   & 6.5       & 6.2   & 4.5   & 2.6       & 5.2   & 6.5           \\
			\midrule qubit (March 2024)    & 11        & 18        & 24    & {\bf 67}  & 70    & 77    & {\bf 81}  & 96    & 108   & 120   \\
			drive freq. [GHz]              & 4.972     & 4.788     & 5.101 & 5.113     & 4.899 & 5.042 & 4.930     & 4.748 & 5.056 & 4.837 \\
			gate error $[10^{-4}]$         & 5.1       & 1.5       & 1.6   & 2.9       & 5.6   & 1.5   & 2.7       & 3.1   & 3.7   & 2.4   \\
			$W^{i}[10^{-5}]$               & -9.6      & -0.8      & -16   & {\bf -40} & -19   & -22   & {\bf -18} & -20   & -3.2  & -1.0  \\
			$\sigma^{i}[10^{-5}]$          & 5.2       & 5.9       & 4.9   & {\bf 3.8} & 4.1   & 5.8   & {\bf 2.6} & 5.1   & 4.0   & 4.5   \\
			$W^{ii}[10^{-5}]$              & -9.8      & -0.9      & -15   & {\bf -40} & -19   & -22   & {\bf
				
				-18}                           & -20       & -3.0      & -1.0                                                                  \\
			$\sigma^{ii}[10^{-5}]$         & 5.2       & 5.9       & 4.9   & {\bf 3.8} & 4.1   & 5.8   & {\bf
				
				2.6}                           & 5.2       & 4.0       & 4.5                                                                   \\
			\bottomrule
		\end{tabular}
		\caption{Experiment results for respective \texttt{ibm\_brisbane} qubits (tests in February and March 2024).
			The data contain drive frequency (inter-level), error of the $S$ gate, the witness $W^{i/ii}$ and the
			standard deviation $\sigma^{i/ii}$.  Indices $i$ and $ii$ of $W$ and $\sigma$ indicate the
				$W$ calculation strategy discussed in section \ref{sec:exp}. Faulty qubits, \cor{with $W$ nonzero beyond 5 standard deviations,} are bolded.}
		\label{data_b}
	\end{table*}

	\begin{table*}
		
		\begin{tabular}{c|*{11}{c}} 
			\toprule \texttt{ibm\_torino} qubit & {\bf 12}  & {\bf 23}  & 33    & 42        & 55    & 60    & 94        & 110   & {\bf 122}   & {\bf 124}   \\
			gate error $[10^{-4}]$         & 1.6       & 1.4       & 1.5   & 1.7       & 2.0   & 2.0   & 1.8       & 1.2   & 1.3  & 1.4   \\
			$W^{i}[10^{-5}]$               & {\bf -166} & {\bf -288} & -50   & -95      & -80   & -19  & -22       & -8.6  & {\bf  -197}   & {\bf -233}    \\
			$\sigma^{i}[10^{-5}]$          &  4 & 	3.4 & 3.6       & 2.4       & 4.5   & 4.5       & 4.3   & 4.5   &  2.9   &  2.4   \\
			$W^{ii}[10^{-5}]$              & {\bf -158} & {\bf -276}  & -50   & -91      & -80   & -19  & -22   & -8.6  &  -194  & {\bf -217}      \\
			$\sigma^{ii}[10^{-5}]$         &  4  & 3.5  & 3.6       & 2.4   & 4.5       & 4.5   & 4.3   & 4.5       & 2.9   &  2.5     \\
			\midrule \texttt{ibm\_pittsburgh} qubit  & {\bf 5}        & 62        & 75    & 86  & 111    & {\bf 115}    & 130  & 145    & {\bf 151}   & 155   \\
			gate error $[10^{-4}]$         & 1.1       & 0.8       & 3.0   & 1.0       & 1.1   & 1.0   & 0.8       & 1.0   & 0.9   & 0.9   \\
			$W^{i}[10^{-5}]$               &{\bf -166}      & -39      & -52   &  -41 & -72   & {\bf -110}   & -52 & -8.2   &{\bf -152}  & -25  \\
			$\sigma^{i}[10^{-5}]$          & 5.5       & 5.5       & 5.5  & 5.5 & 5.5   & 5.5   & 5.4 & 5.6   & 5.4   & 5.4   \\
			$W^{ii}[10^{-5}]$              & {\bf -165}      & -39      & -52   & -40 & -73   & {\bf -110}   &  -52 & -8.1  & {\bf -152}    & -25\\
			$\sigma^{ii}[10^{-5}]$         & 5.5 & 		5.5 & 		5.5   & 5.5 & 5.5   & 5.5   & 5.4  & 5.6&5.4  &  5.4      \\
			\bottomrule
		\end{tabular}
		\caption{Experiment results for respective  IBM Heron devices,  \texttt{ibm\_torino}  and
				\texttt{ibm\_pittsburgh} qubits (test in September 2025). The data contain drive error of the $S$ gate, the witness $W^{i/ii}$ and the
				standard deviation $\sigma^{i/ii}$.  Indices $i$ and $ii$ of $W$ and $\sigma$ indicate the
					$W$ calculation strategy discussed in section \ref{sec:exp}. Faulty qubits, \cor{with $W$ beyond 20 standard deviations,} are bolded. Drive frequencies for  Heron family devices are not disclosed by IBM.}
		\label{data2}
	\end{table*}
	
	\begin{table*}
		\begin{tabular}{c|*{12}{c}}
			\toprule qubit (November 2024)  & 1    & {\bf 3}     & {\bf 5}    & 9    & 13   & {\bf 7}    & {\bf 11}   & {\bf 15}   & 18   & 17   & {\bf 20}    \\
			\midrule gate error $[10^{-3}]$ & 1.1  & 0.9         & 1.6        & 1.2  & 1.0  & 1.3        & 3.0        & 1.3        & 0.6  & 1.6  & 0.6         \\
			$W^{i} [10^{-4}]$               & -2.9 & {\bf -24.7} & {\bf 24.6} & -4.0 & -0.4 & {\bf -6.3} & {\bf -7.5} & {\bf 16.9} & -4.7 & -3.0 & {\bf -17.2} \\
			$\sigma^{i} [10^{-4}]$          & 1.1  & 1.0         & 1.1        & 1.1  & 1.1  & 1.1        & 0.9        & 1.0        & 1.1  & 1.1  & 1.1         \\
			$W^{ii}[10^{-4}]$               & -3.0 & {\bf -24.7} & {\bf 24.6} & -4.0 & -0.4 & {\bf -6.2} & {\bf -7.4} & {\bf 16.9} & -4.7 & -3.1 & {\bf -17.2} \\
			$\sigma^{ii}[10^{-4}]$          & 1.1  & 1.1         & 1.1        & 1.1  & 1.1  & 1.1        & 0.9        & 1.0        & 1.1  & 1.1  & 1.1         \\
			\bottomrule
		\end{tabular}
		\caption{Experiment results for respective  IQM Resonance \texttt{Garnet}  qubits (test in November 2024). The
			data contain error of the $S$ gate, the witness $W^{i/ii}$ and the standard
			deviation $\sigma^{i/ii}$.  Indices $i$ and $ii$ of $W$ and $\sigma$ indicate the
				$W$ calculation strategy discussed in section \ref{sec:exp}. Faulty qubits, \cor{with $W$ nonzero beyond 5 standard deviations,} are bolded. Drive frequencies for IQM Resonance Garnet device are not disclosed by IQM}.
		\label{dataiqm}
	\end{table*}
	
	\subsection*{IQM}
	
\begin{figure}
		\includegraphics[scale=.7]{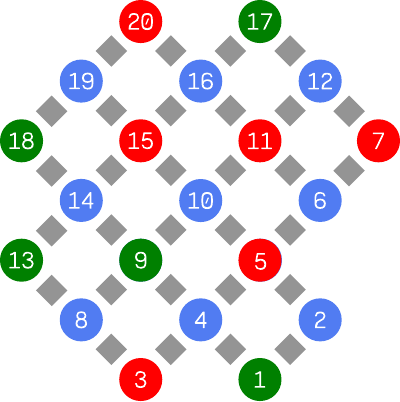}
		\caption{The topology of IQM Resonance \texttt{Garnet}. We highlighted the most faulty
			qubits tested in November 2024 with red. The rest of the qubits, highlighted with green,
			passed the test. Two-qubit CZ gates, unused in the test, connect the qubits.}
		\label{iqmt}
	\end{figure}

	% Setup
	In November 2024, we ran the same test on IQM Resonance - \texttt{Garnet} \cite{iqm}, which contains
	20 transmon qubits. The physical properties of this device are similar those of
	IBM devices. We have chosen 11 qubits so that none is next to the other, see Fig. \ref{iqmt}.
	In this case we have run 13 jobs with $10^4$ shots
	and 5 repetitions.
	
	% Results
	The results turned out to be even worse than for IBM Quantum systems, with a half
	of the tested qubits faulty.  The results are shown in Table \ref{dataiqm}. The largest
	value of the determinant witness $W$ is $\sim 10^{-3}$, which is of the same order as single qubit
	gate error. Similarly to IBM Quantum, the simple leakage is an unlikely explanation.
	
	\begin{figure}
		\includegraphics[scale=.7]{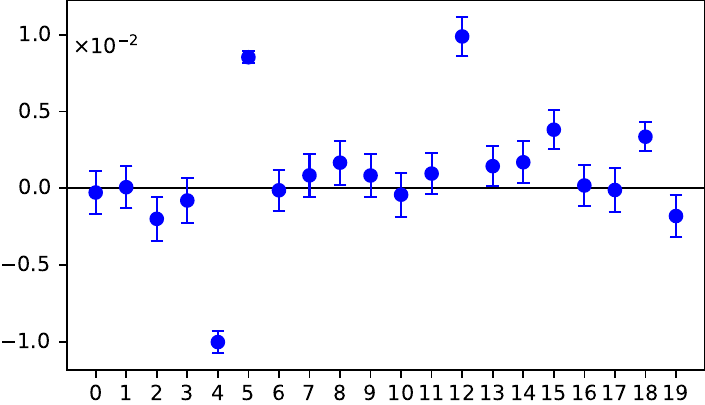}
		\includegraphics[scale=.7]{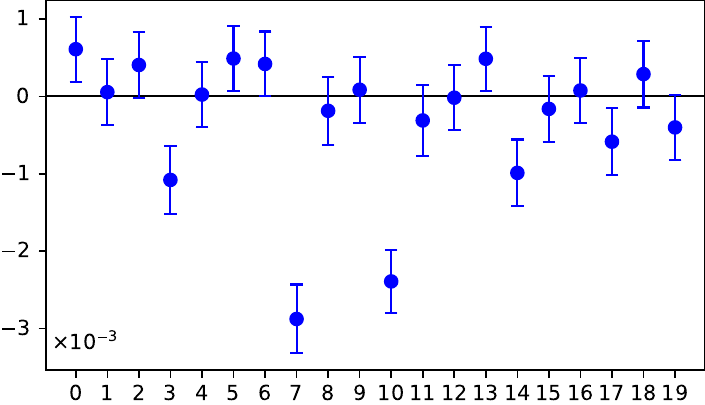}
		\caption{The values of the witness $W$ of individual jobs on IonQ: \texttt{aria-2} (top) and \texttt{forte-enterprise-1} (bottom). Note that
			there are a few values differing from zero by more than 10 standard deviations.}
		\label{ionr}
	\end{figure}
	
	\subsection*{IonQ}
	
	% Setup
	The tests on IonQ \texttt{aria-2} and \texttt{forte-enterprise-1} \cite{ionqd} were a bit different. Due to larger
	single-qubit gate and readout times ($135 \cdot 10^{-6}$~s and $50 \cdot 10^{-6}$~s, respectively),
	but also coherence times $>1$~s, the tests take much longer. For that reason, the results
	become more sensitive to drifts and calibration. We have tested a single qubit 0 (ions
	are coupled all-to-all and are essentially identical), with 20 jobs, each one containing
	5 repetitions and $10^3$ shots (meaning roughly 10 times smaller statistics
	compared to experiments with superconducting devices).
	
	% Results
	A single
	job took more than 1 hour, which is $>6$ times longer compared to 9 minutes on IBM/IQM. The jobs have
	been executed over 6 days in May 2024 on \texttt{aria-2} and over 9 days in May 2025 on \texttt{forte-enterprise-1}. 
The results differed significantly from the ones obtained in the experiments with transmon devices.
	The witness is nonzero beyond 10 standard deviations for a few jobs, indicating sensitivity to calibrations or drifts,
	see Fig. \ref{ionr}.
	
	\section*{Conclusions}
	
	% General results
	A test of the quantum system dimension by parametric rotations reveals
	significant deviations in various qubits of IBM Quantum, IQM Resonance, and
	IonQ devices.
	
	% Detailed overview
	Results form IBM and IQM seem insensitive to calibration drift as both methods of witness value averaging, (i) and (ii),
		lead to similar results. On the other hand, IONQ shows isolated deviations, that indeed may be related to changes in calibration. 
		This needs further investigations. The deviations cannot result from conventional sources like leakage
	to higher excited states of neighbor qubits. This is especially puzzling for
	ion qubits which do not have a natural state to leak to. Moreover, their
	occurrence in a particular period suggests some relation to the unspecified
	calibration change. Large negative deviations found on IBM Heron devices need critical investigation, as they appear consistently across qubits and devices.
	
	% Summary and future work
	Further tests are necessary to identify the deviations'
	origin, to exclude some sophisticated technical causes and more exotic options, e.g.
	many worlds/copies theories \cite{plaga,adp}. We suggest:
	\begin{enumerate}[label=(\roman*)]
		
		\item collecting massive statistics in a relatively short time, which we simply cannot afford,
		
		\item a test on all qubits simultaneously, which is currently impossible due to the
		implementation of classical registers (there is a single multi-outcome counter
		instead of single qubit counters),
		
		\item a test on qubits implemented on  additional platforms (different quantum providers or technology),
		
		\item the use of single-qubit devices (to rule out the cross-talks),
		
		\item applying preparation and measurement on different qubits, with a swap in the middle.
		
	\end{enumerate} 
	Current results confirm that our previously found deviations \cite{epj} were not just a fluke and deserve urgent
	explanation.
	
	\section*{Acknowledgements}
	
	We acknowledge use of the IBM Quantum for this work. The views expressed are those
	of the authors and do not reflect the official policy or position of IBM or the
	IBM Quantum team. We also acknowledge the entire IQM Technology team for their
	support in the development of this work. We also acknowledge the IonQ team for
	their support in the development of this work.
	
	The project is co-financed by the funds of the Polish Ministry of Education and
	Science under the program entitled International Co-Financed Projects. TR gratefully acknowledges the funding support by the program "{}Excellence initiativeresearch university"{} for the AGH University of Krakow as well as the ARTIQ project (UMO-2021/01/2/ST6/00004 and ARTIQ/0004/2021). We thank 	
		Poznań Supercomputing and Networking Center and ICM UW for the access to IBM Quantum Innovation Center. We also thank Bart{\l}omiej
	Zglinicki and Bednorz family for extra access to IBM Quantum.
	
	\section*{Data Availability}
	The data are publicly available at
	doi.org/10.5281/zenodo.17223687
	%\section*{Declarations}
	%\subsection*{Conflict of interest}
	%The authors have no conflicts to disclose.
	\section*{Author contributions statement}
T.R. conceived the experiment, T.R. and T.B. wrote the codes, A.B. conducted the experiments, T.R. and A.B. analyzed the results. A.B. and J.B. wrote the manuscript, J.T. improved the manuscript. All authors reviewed the manuscript.

\end{document}